\documentstyle[12pt,epsfig]{article} 
 \newcommand {\bi} {\bibitem}
 \newcommand {\be} {\begin{equation}}
\newcommand {\bea} {\begin{eqnarray} \nonumber }
\newcommand {\ee} {\end{equation}}
\newcommand {\eea} {\end{eqnarray}}
 \newcommand {\eps} {\epsilon}

\newcommand {\la} {\langle}
\newcommand {\ra} {\rangle}
 \newcommand {\al} {\alpha}
\def\(({\left(}
\def\)){\right)}
\def\[[{\left[}
\def\]]{\right]}
\newcommand{\nn}{\nonumber}

\def \pt{{\tilde p}}

\begin{document}


\title{How to compute the thermodynamics of a glass \\using a cloned liquid }
\author{ Marc M\'ezard \\
{\it Institute for Theoretical Physics}\\{\it University of California Santa Barbara,
 CA 93106-4030, USA}\\ 
{ \it and}\\
{\it Physique Th\'eorique-ENS, CNRS, France}}
\date{\today}

\maketitle
\abstract{The recently proposed strategy for studying the equilibrium
thermodynamics of the glass phase
using a molecular liquid is reviewed and tested  in details 
on the solvable case of the $p$-spin model.
We derive the general phase diagram, and confirm the validity of this procedure.
We point out the efficacy of a system of two weakly coupled copies in order to
identify the glass transition, and the necessity to study a system with $m<1$
copies ('clones') of the original problem in order to derive the thermodynamic properties
of the glass phase.
}

\section{Introduction}
It has been proposed recently that the thermodynamic properties of the 
glass phase can be deduced from those of the corresponding
liquid phase, computed above the glass transition but for a 
`replicated' liquid involving $m$ species with $m<1$,
forming then a {\it molecular bound state} \cite{MPglass}. This scheme is appealing
in that it provides a general strategy for computing the 
equilibrium properties of the glass phase\cite{rem_eq}. This whole strategy
is supposed to be valid whenever the thermodynamic transition is due to an
'entropy crisis' scenario \cite{kauzman,AdGibbs,Derrida,GrossMez,KiThiWo,glass_revue}.
This basically means
that the configurational entropy defined
 as the logarithm of the number of accessible metastable states of the system, which is
extensive above the transition temperature, becomes zero below the transition temperature.

The archetypes of such systems, and the only ones for which one can get a full control
so far, are provided by the so-called $p$-spin models,
some infinite range spin glasses involving $p>2$ spin interactions,
which have been already much studied. In this
paper we want to test the 
general strategy of \cite{MPglass} on these exactly solvable
cases. We show explicitely how the properties of the glass phase 
in these systems can be deduced from
that of the replicated liquid, and we point out some subtleties of the phase
diagram of this replicated liquid, which should be kept in mind when one
deals with the real systems of interacting particles. 

This paper complements the 
 work by Franz and Parisi \cite{FraPar} which performed a similar
study of the phase diagram of two coupled systems when one of the system is thermalised
in the low temperature phase. Here we concentrate instead on the case where we use
an arbitrary number $m$ of coupled systems which  are all identical, as used 
in \cite{MPglass} (The case $m=2$ was alluded to in \cite{FraPar}).

\section {The basic strategy}

\subsection{Identifying the glass phase}
Let us assume that there exists a glass phase in some system.
How can one distinguish the glass phase from the liquid one, in the 
framework of equilibrium statistical mechanics? A simple answer, developed
initially
in the spin glass context \cite{parisi83,Toulouse,carparsour,FraParVir}, is the following. 
Take two copies, with a small
attraction between particles in the two copies, of strength $\eps$.
The attraction is short range (range less or equal to the typical interparticle
distance), and it couples all particles of system 1 to all
particles of system 2. Look at the  correlations
between the two copies  when $\eps \to 0$ (limit taken after the thermodynamic limit).
If there  remain some  correlations in this double
limit, this means that the system is in a solid phase (a glass or a crystal); otherwise
it is in the liquid phase. This general technique is a very robust one
which is able to identify the glass phase in very diverse systems,
ranging from spin glasses to directed polymers and to structural glasses
\cite{parisi83,Toulouse,carparsour,mez_pol,MPglass}.

In structural glasses this transition is associated with a spontaneous breaking 
of a translation symmetry. For $\eps=0$, the
 Hamiltonian  is invariant by the relative global
translation of the particles in the second system relative to the particles in
the first one (assuming periodic boundary conditions). This symmetry is spontaneously broken
in the solid phase.
Therefore one expects that there is no way in which one can connect continuously the
liquid phase to the glass phase without encountering a phase transition.
We shall check this property  explicitely on the case of the
$p$-spin mean field spin glass: we shall work out the phase diagram in the plane $\eps,T$,
showing the existence of a glass transition temperature at all $\eps$. 

\subsection{Nature of the glass transition in the discontinuous case} 
This identification of the transition from the study of two coupled system
 does not allow to reach the properties of the low temperature
phase from a study of the high temperature one, precisely because of the 
existence of a phase
transition. It turns out that, for a specific class of systems,
 there exists a more subtle generalization of this approach
which uses an arbitrary number $m$ of coupled systems and allows to study
the glass phase. These are the systems for which the glass transition is of
a discontinuous type, called `one step replica symmetry breaking' (one step rsb) in the
spin glass language. Basically this situation is encountered whenever there exist
some low-lying states of the system which are well separated in phase space
and have independent free energies \cite{MPV_randomfree}, a situation which
has been argued to give a good description of the structural glass phase
\cite{parisi_new_ideas}.
 More precisely, the extensive part of their free
energies are equal, but the fluctuating, non-extensive part of their free
energies are not correlated from state to state. One may expect that such systems
will be generic in the sense that this situation corresponds to a universality
class for extreme event statistics \cite{BouMez_extr}. Proving whether a given physical
system actually falls in such a class may be quite difficult, but a useful
approach, adopted in \cite{MPglass}, consists in
assuming that it does, computing the
properties of the system within such an assumption, and comparing them to experimental and
numerical data. The success of this strategy, and the number of experimental
obsevations that it explains and relates, have added confidence to the
validity of the old conjecture
\cite{KiThiWo} according to which fragile structural glasses \cite{glass_revue}
fall into this category 
\cite{parisi_new_ideas,MPglass}.

In this paper we shall concentrate on the mean field spin glasses with $p$-spin interactions
($p>2$)
which have been shown to belong to this class of systems, both for Ising
spins and for spherical spins
\cite{GrossMez,KiThiWo,crisanti,kurparvir} 
(Several other sytems have been proven to belong to this class, including 
some spin glasses without quenched disorder \cite{nodis}).
We shall briefly review some of their
properties which are crucial for understanding the nature of their glass transition
and how the glass phase can be studied.
These systems possess an infinity
of metastable states, which we shall label by an index
$\alpha$. Each state is characterized \cite{MPV}
by the value of the local magnetization on each site, $m_i^\alpha=\la s_i\ra_\alpha$.
The states, and their free energies,
 can be defined as solutions of TAP equations.
A very important quantity is
 the number $\cal N$ of such metastable states with a given free energy density $f$. It turns out
 that, in some range of free energies,
  this number is exponentially large in the number of degrees of freedom. One can thus define a
  corresponding entropy, called the configurational entropy in the glass litterature
  \cite{glass_revue}
  and the complexity in the spin glass litterature \cite{Palmer,MPSTV,GrossMez}, defined
  by 
\be
{\cal N}(f,T,N) \approx \exp(NS_c(f,T)),\label{CON}
\ee
The number of metastable states  vanishes outside of the region 
$f_{min}(T)<f<f_{max}(T)$, and the configurational entropy $S_c (f,T)$
 goes to zero at $f_{min}(T)$.
 
 Let us discuss the properties of such a  system at thermal equilibrium. We
 call $f_\alpha$ the free energy density of state $\alpha$. At low enough temperatures
 the total free-energy-density of the system ($f_{S}$) can be 
well approximated by the sum of the contributions to the free energy of each particular minimum:
\be
Z\equiv \exp(-\beta N f_{S}) \simeq\sum_\al \exp(-\beta N f_\al).
\ee
For large values of $N$ we can write
\be
\exp(-N \beta f_{S}) \approx \int_{f_{min}}^{f_{max}} df \exp (-N(\beta f- S_c(f,T)).\label{SUM}
\ee
Using the saddle point method we find that
\be
f_{S}=\min_f\Phi(f) \equiv  f^* -  T S_c(f^*,T),
\ee
where
\be
\Phi(f)\equiv f - T S_c(f,T).
\ee
This formula is quite similar to the usual formula for the free energy, i.e.  
$f=\min_{E} ( E - T S(E))$, where $S(E)$ is the entropy density as 
a function of the energy density.
The main difference is the fact that the total entropy of the 
system has been decomposed into the
contribution due to small fluctuations around a given configuration (this piece has
been included into $f$), and the contribution due to the existence of a large number of
locally stable configurations, the configurational entropy.

\begin{figure}
\centerline{\hbox{
\epsfig{figure=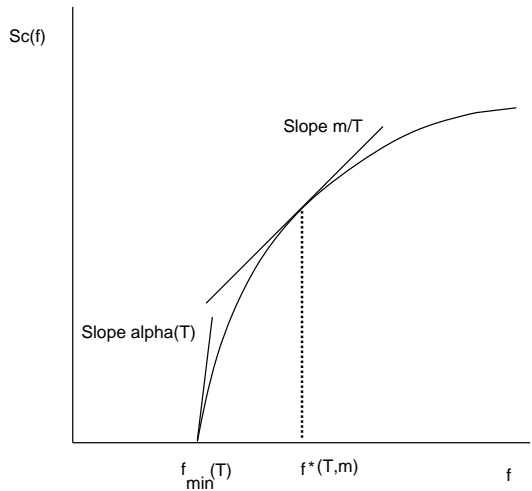,width=7 cm,angle=0}
}}
\caption{Qualitative shape of the configurational entropy versus free energy. The
whole curve depends on the temperature. The
saddle point which dominates the partition function,
for $m$ correlated clones, is the point $f^*$ such
that the slope of the curve equals $m/T$ (for the usual, uncloned, system, $m=1$). 
If the temperature is small enough the saddle point sticks
to the minimum $f=f_{min}$ and the sytem is in its glass phase. }
\label{confentrop}
\end{figure}

Calling  $f^*$ the 
value of $f$ which minimizes $\Phi(f)$, we have two possibilities:
\begin{itemize}
\item
The minimum of $\Phi$ lies inside the interval and it can be found as the solution 
of the equation $\beta=\partial S_c/\partial f$.  In this case we have
\be
 \Phi= f^* - T S_c^*, \ \ \ S_c^*=S_c(f^*,T).
\ee
The system may stay in one of the many possible minima.  The number of accessible minima 
 is $\exp(N S_c^*)$ .  The entropy of the system is thus 
the 
sum of the entropy of a typical minimum and of $S_c^*$, which is the contribution to the entropy 
coming from the exponentially large number of metastable configurations.

\item
The minimum of $\Phi$ is at the extreme value of  the range of variability of 
$f$: it sticks at $f^*=f_{min}$ and the total free energy is
$\Phi=f_{min}$.  In this case the contribution of the configurational entropy to 
the 
free energy is zero.  The different states which contribute to the free energy have a difference in 
free energy density which is of order $N^{-1}$ (a difference in total free energy of order 1),
and these free energy differences are uncorrelated and exponentially distributed
\cite{MPV_randomfree,MPV}, corresponding to Gumbel's distribution of extremes \cite{BouMez_extr}.

\end{itemize} 
 
 Therefore the glass transition of this system is a thermodynamic transition associated
 to what is sometimes called an `entropy crisis': the entropy associated with the
 number of metastable states which 
 contribute to the partition function vanishes below the transition temperature.
 The transition is of second order from the thermodynamic point of view,
 having a continuous entropy and a jump in specific heat, but it
 is associated with a discontinuity of the order parameter \cite{GrossMez}. We
 call such a transition a 'discontinuous glass transition'.
 
 \subsection{Cloning} 
 \label{constrained}
 Clearly the equilibrium thermodynamics of this class of glasses is fully described by the knowledge
 of the configurational entropy as a function of temperature and free-energy-density.
 Monasson \cite{remi} proposed the following method to compute this function.
 l
 Let us consider
 $m$ copies of the system which are constrained to stay in the same minimum. We shall discuss below
 how one can achieve this constraint, but let us first discuss the physics of this 
 constrained system.
   Its partition function  is basically the Laplace transform of the number of states:
\be
Z_{m} = \int_{f_m}^{f_M} df \ e^{-N [m \beta f- S_c(f,T)]} \ .
\ee
The   free
energy per spin,
\be
\phi(m,T)= -{1 \over \beta mN} \log Z_m \approx
\min_{f}( \ f - {T \over m} S_c(f,T)) \ ,
\label{free_rep}
\ee
allows to compute  the configurational entropy $S_c(f,T) $. 
Because of the concavity
of  the $S_c$ versus $f$ curve, its largest possible
slope is 
\be
\alpha(T) \equiv {\partial S_c \over \partial f}(f_{min},T) \ .
\ee
Performing the minimisation in (\ref{free_rep}), 
we find that the states which contribute to the partition function
 have a free energy density $f^*(m,T)$
given either by 
\be
{\partial S_c(f,T) \over \partial f} (f^*,T)= {m \over T}
\ee
if $m$ is small enough (so that the slope
at the origin, $\alpha(T)$, be
larger than $m/T$), or otherwise
they are just the low lying states with $f^*=f_{min}$ (see fig. \ref{confentrop}). 
Therefore, varying $m$
at a fixed temperature $T<T_c$, we have a phase transition at a critical value
of $m=m_c(T)$ which is {\it smaller than one}. 

\begin{figure}
\centerline{\hbox{
\epsfig{figure=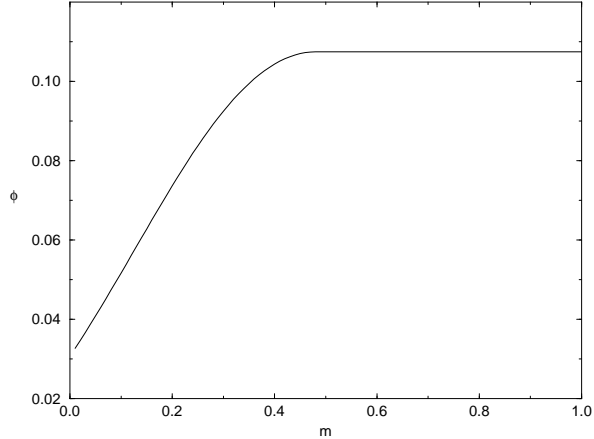,width=7 cm,angle=-90}
}}
\caption{General shape of the free energy per particle $\phi(m,T)$ of the cloned system,
versus the number of clones $m$, at temperatures below the glass transition. The critical
value $m=m_c(T)$ is identified as the first point where 
 the slope $\partial \phi /\partial m$ vanishes when one increases $m$.}
\label{phidem}
\end{figure}

The general shape of the free energy $\phi(m,T)$ is the one shown in fig.\ref{phidem}.
For $m>m_c(T)$ the free energy is constant $\phi(m,T)=f_{min}(T)$. For $m<m_c(T)$
this function allows to compute the configurational entropy through the
implicit equation:
\bea
{\partial [m \phi(m,T)] \over \partial m}&=& f\\
{m^{2} \over T} {\partial \phi(m,T) \over \partial m}&=& S_c  \ .
\label{conf_entr_gene}
\eea
Increasing $m$, the transition point $m_c$ is determined by the vanishing
of the slope ${\partial \phi(m,T) \over \partial m}$.

The following two remarks, which are trivial consequences of the simple nature
of the phase transition in these discontinuous sytems, are crucial:
\begin{itemize}
\item
For $m<m_c(T)$, the system is in its high temperature phase, even if $T<T_c$. 
The free energy $\phi(m,T)$ is
that of the high temperature phase of a system of $m$ constrained copies, a liquid
made of molecules built up from $m$ atoms. It can thus be
evaluated by any of the methods giving a good description of the high 
temperature phase, including series expansion, simulations etc...
\item
The transition at $m=m_c(T)$ is such that the free energy is constant for $m>m_c(T)$ (this
is a consequence of the transition being due to the sticking of the 
saddle point in $f$ at the lowest value $f_{min}$). Therefore
the free energy in the glass phase at a temperature $T<T_c$ is given by
$F(T)=\phi(m=1,T)=\phi(m_c(T),T)$. But the free energy is continuous through the
transition line, and therefore $\phi(m_c(T),T)$ is the free energy of the 
 high temperature, 'liquid', phase.
 \end{itemize}
 
 This is the method which has been used in \cite{MPglass} to compute the thermodynamics of
the glass.  It involves dealing with a cloned system made up of $m$ identical clones
constrained to be in the same state, and to do this for any $m$, in such a way as 
to be able to
analytically continue the results to $m<1$. This continuation bears some similarities
to, but is different from, the one used in the replica method in order 
to compute quenched averages. As we shall also need to use these more standard
replicas, we shall call the  $m$ copies
introduced here and in \cite{remi,MPglass} some 'clones' of the original
problem (we keep the word copies for the case when $m$ is an integer
larger or equal to one). One can introduce the clones even for systems which have no
quenched in disorder, like structural glasses. The reader will have noticed
that the above discussion is a general one which does not rely on the details of
the $p$-spin mean field models; in fact the  procedure of computing the free energy
of the cloned and constrained system, $\phi(m)$, in the liquid phase, can be
used also in some non mean field systems \cite{MPglass}, 
in which it is more difficult to come
up with a definition of
the metastable states, since these involve fixing some state-dependent
boundary conditions in a self consistent way.
The replicas introduced in disordered systems, which are used to handle the average over
quenched disorder, can be introduced on top of the clones if needed:
in the case of systems with quenched disorder, clones can be replicated! As we shall see,
the computation with the fixed number $m$ of coupled
clones in the high temperature phase is equivalent for disordered systems to a
 `one step rsb' computation. This has been first found in
 some computations of neural networks
 \cite{oKane} and is expected on general grounds \cite{remi}.
 
 We would like to also 
 comment on the alternative method which  has
  been proposed independently for the study of the glass phase \cite{pot}. It also
  introduces some clones of the system, but it uses one reference system which is
  supposed to be thermalised in the glass phase, and then studies the free
  energy of some other systems, constrained to be at a fixed distance of the reference one.
  This is a powerful method which has the advantage 
  of allowing to study of metastable states in the high temperature
  phase, as has been demonstrated in \cite{FraPar,CarFraPar}.
  It has the drawback of requiring to have one system thermalised in the
  glass phase, which is not easy to achieve. 
  The strategy of cloning discussed here, which has been used
  in \cite{remi,MPglass} has the advantage of
  involving only computations in the liquid phase, but its 
  scope is probably  limited to the study of equilibrium thermodynamics.
 
 \subsection{Two coupled copies}
We have seen in the previous section how a set of $m$ clones constrained to stay in the same
state can allow for the computation of thermodynamical properties of glasses. The 
question we address now is how to achieve this constraint. In order to try and constrain the
$m$ clones to be in the same state, we shall introduce an attractive coupling between the 
various clones. For spin systems, calling $s_i^a$ the spin at site $i$ 
in clone $a$, this coupling could be for instance: 
\be
- \eps \sum_{1 \le a<b \le m} \sum_{i=1}^N s_i^a s_i^b \ .
\ee
 The detailed structure of this coupling is to some extent irrelevant, but it
 should satisfy the following constraints: it is extensive, in that it contributes
 to the energy at order $N$, and it is proportional to $\eps$. What we shall be 
 mostly interested in is taking the double limit $\lim_{\eps \to 0} \lim_{N \to \infty}$.
 But it will also turn out to be useful to study the phase diagram of the coupled
 system for general $\eps$. This we shall do in the next section for the $p$-spin models,
 but let us first try to get a qualitative understanding of the phase diagram
 in general, based only on the existence of the configurational entropy. The various parameters
 in our problem are the number of clones $m$, the temperature $T$, and the
 coupling between clones $\eps$.

Let us first consider the case  of two coupled copies, $m=2$, in the limit of
small $\eps$.
 There can be several contributions
to the partition function: 

- When the two systems fall into the same state, we have: 
\be
Z_{=} \simeq \int_{f_{min}(T)} df \ e^{N[S_c(f,T)-2f/T + \eps q_1/T]} \equiv e^{-2 N \phi_=/T}
\label{Zeq}
\ee
where $q_1$ is the typical self overlap of a state.

- When they are in different states, we have:
\be
Z_{\ne} \simeq \int_{f_{min}(T)} df \  e^{N[2 S_c(f,T)-2f/T]}\equiv e^{- 2 N \phi_{\ne}/T}
\label{Zdiff}
\ee

There exist several temperature regimes:

- If $T<T_c(0)$, both contributions are dominated by the boundary $f=f_{min}$,
giving $\phi_==f_{min}-\eps q1/2$ and $\phi_{\ne}= f_{min}$. Therefore
for any positive $\eps$, we have $\phi_=<\phi_{\ne}$ and
 the leading contribution is due to systems
falling into the same state. The system is in a correlated phase because the overlap between
the two copies is non zero, and it is in the glass phase because the free energy
sticks at $f_{min}$.

- When $T_c(0)<T<T_2$, where $T_2$ is the temperature at which $\alpha(T_2)=2/T_2$, we have
 $\alpha(T)<2/T$. Thus the contribution $Z_=$ is dominated by the boundary and
  $\phi_= =  f_{min} -\eps q_1/2$. The contribution $Z_{\ne} $ is dominated by 
a non trivial sadle point $f^*$
 such that: $(\partial S_c /\partial f)(f^*) =1/T$. The corresponding
 free energy is $ \phi_{\ne}= f^*-T S_c(f^*)$. Depending on the value of 
 $\eps$, the minimal free energy will be either $\phi_=$, in which case the system is in
 a correlated glass phase, or $\phi_{\ne}$, in which case the system is in an uncorrelated 
 liquid phase.
  Because of the concavity of the $S_c(f)$
curve, $ f^*-T S_c(f^*) < f_{min}$ and therefore the 'liquid' saddle point
dominates at $\eps$ small enough. The critical value of $\eps$ locating the
transition to the
glass phase, with the two systems sticking together, is
\be
\eps_c=(f_{min}-f^*+T S_c(f^*))/q_1
\ee
This whole discussion holds only at small $\eps$, and in this 
region one can locate the behaviour of the transition curve
in the neighborhood of the point $T=T_c, \eps=0$, by using the
following expansion of the configurational entropy at $f>f_{min}$:
\be
S_c(f)=\alpha(T) (f-f_{min}(T)) - {\gamma(T)  \over 2} (f-f_{min}(T))^2 + O\(((f-f_{min}(T))^3\)) \ .
\ee
We have then:
\be
f^*\simeq f_{min}+{1 \over \gamma(T)} \(( \alpha(T)-1/T\))
\ee
In the neighborhood of the point $T=T_c, \eps=0$, the
generic behaviour is that $\alpha(T)-1/T$ is
linear in $T-T_c(0)$. Writing $\alpha(T)-1/T \sim g (T-T_c(0))$,  we get finally:
\be
T_c(\eps)-T_c(0)\sim \sqrt{{\gamma q_1 \over T_c(0) g^2}} \ \sqrt\eps
\ee

This discussion suggest that, in the limit of small coupling ($\eps \to 0$),
the system of two coupled copies will have a transition to the glass state
at a temperature which depends continuously on $\eps$, with a square root
singularity. This transition is
clearly identified by a first order jump in the correlation between the
two systems, defined here as the overlap.
 Such a system with
two copies  may thus be a good candidate for the numerical determination of $T_c(0)$,
as already discussed in \cite{FraPar}.

Let us also notice that there might in principle exist another phase of the 
system, with the two copies correlated, but being in the liquid phase. (This
is the phase of the molecular liquid). This can exist whenever 
 $T_2<T$:
The contribution $Z_{\ne} $ is then dominated by 
a non trivial sadlle point $f^*_{\ne}$
 such that: $(\partial S_c /\partial f)\((f^*_{\ne} \)) =1/T$. This is also the case
 of the contribution $Z_=$, with a saddle point $f^*_{=}$
 such that: $(\partial S_c /\partial f)(f^*_=) =2/T$.
Concavity implies  that $ \phi_{\ne}<\phi_=$ at small enough $\eps$.
Therefore the  system is in its uncorrelated liquid phase for $\eps$
small enough, at all temperatures $T>T_c$: in a system with two coupled copies
one does not
see the correlated liquid phase at small $\eps$. 
If one could use  these small $\eps$ computations to 
induce some conclusions for finite $\eps$, they would hint at a transition,
in the liquid phase, between the uncorrelated one at small $\eps$ and the correlated
one at larger $\eps$. 

In fact for very large $\eps$ the two coupled systems will be in the same configuration, 
giving a partition function $Z=\int \exp(-2 \beta H)$. Therefore
this system will have a usual glass transition at a static temperature:
\be
\lim_{\eps \to \infty} T_c(\eps) =2 T_c(0)
\ee
and this transition will be a discontinuous glass transition, therefore
it will have no entropy jump.

The explicit computations in the $p$-spin models will confirm the various 
aspects of the phase diagram of the two copies system that we have 
predicted here. We expect that these are general features which will be found in all
the glass phases.

\subsection{Coupled clones}
Let us now study the main aspects of the phase diagram that one can expect when there is
a general number, $m$, of clones.
As we have seen above, the system can be, for small $\eps$,
in any of the following three
possible phases:

\begin{itemize}
\item
A correlated glass phase in which  the free energy is $\phi=f_{min}- (m-1) \eps q_1/2$.
\item
A correlated liquid phase in which  the free energy is the saddle point of
$\phi= \ f - (T/m) S_c(f,T)- 
(m-1) \eps q_1/2$, and the extremum value $f^*$ is greater than $f_{min}$.
\item
An uncorrelated liquid phase in which
 the free energy is the saddle point of $\phi=  \ f - T S_c(f,T)$ and the extremum 
 value $f^*$ is greater than $f_{min}$
\end{itemize} 

One has to distinguish two regions in $m$ space. 

If $m>1$, the discussion is very similar
to the one we did above for the $m=2$ phase, and we expect that at small $\eps$
the system will have a transition from an uncorrelated liquid phase at $T>T_c$ to a correlated
glass phase at $T<T_c$.

If $m<1$ the situation is different, basically because the interesting part
of the phase diagram now lies in the $T<T_c$ region. 
When $T<T_c$, 
the uncorrelated liquid cannot exist because $\alpha(T)>1/T$. Therefore one expects a
transition from the correlated liquid to the correlated glass, at a temperature such that
$\alpha(T)=m/T$. This is precisely the transition which  was discussed in the
previous section. So our conclusion is that the $\eps$ coupling, for small $\eps$
and $T<T_c$, is able to polarize the system in the same state 
provided $m<1$, and thus it realizes
the constrained system that was discussed in sect. \ref{constrained} and allows
for the computation of the glass thermodynamics from liquid computations.

\section{Explicit computations with the  $p$-spin model}
We now move on to some explicit computations of the phase diagram of the
cloned $p$-spin model.
\subsection{The usual replica solution}
We work with a spherical spin glass model \cite{crisanti} involving $N$ spins
$S_i$ with the spherical constraint $\sum_i S_i^2=N$ and an energy given by:
\be
H=-\sum_{i_1<...<i_p} J_{i_1...i_p} S_{i_1}...S_{i_p}
\ee
where the random couplings $J_{i_1...i_p}$ are independent gaussian random variables
with zero mean and variance $p!/(2 N^{p-1})$. In the explicit computations below,
and for all the numerical results,
we shall use the value $p=4$; however all the systems with $p>2$ have the same
qualitative phase diagram. We shall not reproduce  here the various
techniques which have already been used to study this system, but just
state some results of interest for our discussion, referring the reader to the original
litterature \cite{crisanti,crisomtap,kurparvir}. The replica solution shows the
existence of a transition from a replica symmetric phase at $T>T_c$ to
a phase with one step rsb. The breaking involves a parametrization
of the overlap matrix between $n$ replicas, $Q_{ab}=(1/N) \sum_i
 \la S_i^a S_i^b \ra$, which
has a one step rsb Parisi structure \cite{MPV} with a diagonal element equal to one,
the elements inside diagonal blocks of size $x$ equal to $q_1$, and 
all the other elements equal to zero.
The free energy is
\be
F=-{\beta \over 4}[1-(1-x) q_1^p] + {1-x \over 2 \beta x} \log(1-q_1)
-{1 \over 2 \beta x} \log[1-(1-x) q_1]
\label{free_noclone}
\ee
and should be maximised over $q_1$ and over $x$ in the interval $[0,1]$.
 
The high temperature phase is obtained when $x\to 1$. In the high temperature phase 
$q_1=0$ and the free energy is just $-\beta/2$. 

The parameter $q_1$ is interpreted as the Edwards-Anderson (EA) order parameter,
characteristic of the degree of spontaneous ordering inside each state:
$q_1=(1/N) \sum_i \la S_i \ra_\alpha^2$. The low temperature phase has
an EA order parameter $q_1 \ne 0$, and $q_1$ jumps discontinuously at
$T_c$, although the transition is thermodynamically of second order, with
a continuous entropy and energy (this is possible because $x \to 1$ at the transition
\cite{GrossMez}). The parameter $x$ is characteristic of  the distribution of
free energies of the low lying states \cite{MPV_randomfree,MPV}.

\begin{figure}
\centerline{\hbox{
\epsfig{figure=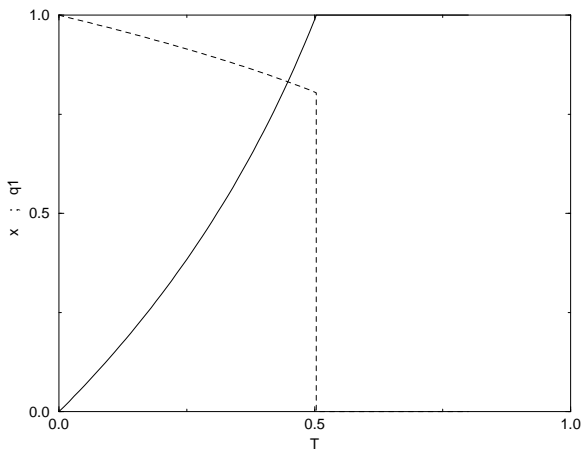,width=7 cm,angle=-90}
}}
\caption{Replica solution of the p=4 model at equilibrium. We plot the Edward-Anderson order
parameter $q_1$ (dashed line) and the one step rsb
parameter $x$ (full line) versus the temperature. The transition is at the 
temperature $T=.503$}
\label{fig1}
\end{figure}

The result for $p=4$ is summarised in fig.\ref{fig1}, which plots the
value of the breakpoint parameter $x$ and the EA parameter
versus temperature. The transition temperature,
identified from $x=1$, is $T_c=.503$. The solution with $x >1$ is interesting
for the study of metastable states and barriers \cite{kurparvir}, 
but here we stay within the 
framework of the equilibrium measure where it does not play any role.

\subsection{Two coupled copies: exact solution and phase diagram}
We now turn to a coupled system with $m=2$ identical copies of the original problem, two
sets of spins $S_i$ and $S_i'$, characterized by the 
energy:
\be
H=-\sum_{i_1<...<i_p} J_{i_1...i_p} \((S_{i_1}...S_{i_p}'+S_{i_1}'...S_{i_p}' \))
 -\eps \sum_i S_i S_i'
\ee
The thermodynamics of a cloned $p$-spin model with some coupling between the clones
has already been studied in \cite{kurparvir} in order to understand the structure of 
metastable states and barriers.
 We shall borrow from this paper the form
of the free energy
 of the coupled system, computed with the replica method (replicated clones),
 within the one step rsb
 formalism. 
 The replica formalism introduces $n$ replicas of each of the two spin systems.
 The free energy is expressed in terms of a self-overlap matrix between
 $n$ different replicas of the spins $S_i$ , $Q_{ab}=(1/N) \sum_i
 \la S_i^a S_i^b \ra= (1/N) \sum_i
 \la S_i'^a S_i'^b \ra$. It also involves the correlations between the replicas of
 the two sets of spins, $P_{ab}=(1/N) \sum_i
 \la S_i^a S_i'^b \ra$. Within the one step rsb solution, which ws shown 
 to be correct in \cite{kurparvir}, the matrix $Q$ is parametrized exactly
 as in the previous case by the value of its elements, $q_1$, inside the diagonal
 blocks of size $m$, while the matrix $P$ is parametrized 
  by the value of its elements, $p_1$, inside the diagonal
 blocks of size $m$, and the value $\tilde p$ of its diagonal elements 
 (all other elements being 
 $0$). In terms of these order parameters, the free energy is given by:
 
\bea
F= &-&\beta[f(1)+f(\pt)-(1-x)(f(q_1)+f(p_1))]
\\ \nn
&-& {1 \over 2 \beta} {1 \over x} [\log(1-\pt+(1-x)(p_1-q_1))+\log(1+\pt-(1-x)(p_1+q_1))]
\\
&+&{1 \over 2 \beta} {1-x \over x} [\log(1-\pt+p_1-q_1)+\log(1+\pt-(p_1+q_1))] -\eps \pt
\eea
Where $f(r)\equiv r^p/2$.
We have two real copies, so the free energy must be minimized with respect to $\pt$.
But we have $n \to 0$ replicas to handle the quenched average, so 
the free energy must be maximized with respect to $p_1,q_1,x$ \cite{MPV}.
 
The interpretation of the various order parameters is as follows.
The system may freeze into a correlated glass phase. This will be signalled by the 
Edwards-Anderson-like  order parameters:
\be
q_1= {1 \over N} \sum_i \la S_i\ra _\alpha \la S_i' \ra _\alpha \ , \ 
p_1= {1 \over N} \sum_i \la S_i\ra _\alpha \la S_i \ra _\alpha
\ee
On the other hand, for any finite $\eps$, and both in the high and low temperature
phase, the degree of correlations between
the two systems is measured by:
\be
\pt= {1 \over N} \sum_i \la S_i S_i'\ra 
\ee

\begin{figure}
\centerline{\hbox{
\epsfig{figure=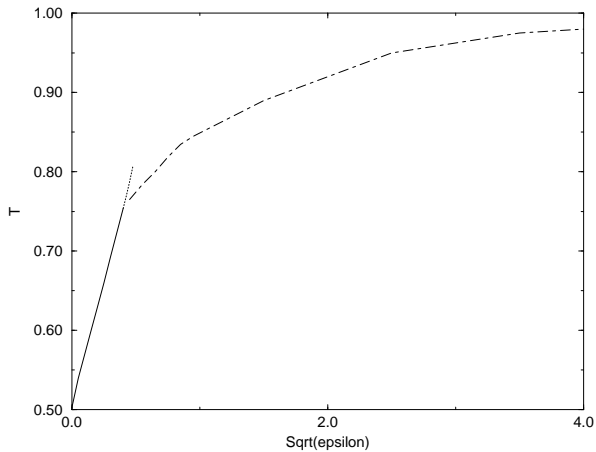,width=7 cm,angle=-90}
}}
\caption{Phase diagram in the plane $\sqrt\eps$, $T$, for a system of two coupled
copies of a $p$-spin model with coupling $\eps$, and $p=4$. 
Full and dashed-dotted lines: the glass transition temperature $T_c$, identified by a jump in the
Edwards-Anderson parameter, which is zero in the high temperature
phase, and jumps discontinuously to a positive value in the low temperature phase.
The full line is a first order transition with a latent heat and a jump in the 
correlation (see text), while the dashed-dotted line is a 'discontinuous spin glass'
transition with no latent heat, no jump in the correlation, but a jump in the EA order
parameters.
The dotted line is a first order transition line, ending at a second order
critical point, separating two regions of the liquid with
different correlations between the two copies.}
\label{fig9}
\end{figure}

The extremisation of the free energy at $p=4$ 
and finite $\eps$, performed by a simple grid search followed by a minimisation procedure,
 gives the
 phase diagram plotted in fig. \ref{fig9}.

 Above the full and dashed-dotted  lines, the system is
 in a liquid phase, where the EA order parameters are $q_1=p_1=0$. The 
 correlation $\pt$ is nonzero, but it goes to zero in the limit of $\eps \to 0$.
  Below the full and the dashed-dotted lines the system is in the glass phase, with $q_1=p_1>0$.
  The
 correlation $\pt$ is nonzero, and it remains finite when $\eps \to 0$. Therefore
 the $\eps \to 0$ transition
  is between  an uncorrelated liquid phase at high temperature and a correlated
  glass phase at low temperatures. It  occurs at the temperature $T_c(0)=.503$,
  with some small $\eps$ corrections which are of order $\sqrt\eps$, as anticipated
  in our general discussion of sect. \ref{constrained}.
   This glass transition  is a first order
  transition on the portion with small enough  $\eps $ (full line).
   On this portion the various order parameters 
  $q_1,p_1, \pt, x$ jump
  discontinuously at the transition. In particular the correlation parameter $\pt$ is
  discontinuous on this transition, which should make it relatively
  easy to detect numerically (letting aside metastability effects),
   and there is also a latent heat. The  glass transition at
   larger $\eps$ (dashed-dotted line) is of the second order type
   from the thermodynamic point of view. It has a jump in the EA
   order parameter $q_1=p_1$, but the correlation parameter $\pt$ and the 
   entropy are continuous at the transition; it is thus a transition which is exactly of the 
   'discontinuous spin glass type', as it is known to occur both for the single system
   $\eps=0$, as well as for the infinitely coupled system $\eps \to \infty$. 

\begin{figure}
\centerline{\hbox{
\epsfig{figure=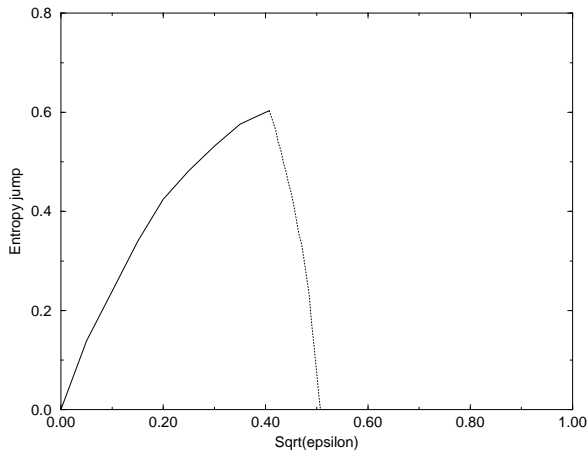,width=7 cm,angle=-90}
}}
\caption{Discontinuity of the entropy along the two lines of first order
transition (full line and dotted line) in fig. \ref{fig9}. The transition
becomes of second order (continuous entropy) at the two end points: the end point
at $\eps=0$ which is the usual glass transition, and the critical point for the decorrelation
transition in the liquid phase. The entropy jump is continuous along the
triple point, but its derivative is discontinuous. }
\label{entrop_jump}
\end{figure}

     The
   first order transition line  goes on into the liquid phase, where there is a portion
   (dotted line) which is a first order transition line. This transition has
   {\it nothing to do} with the glass transition: on both sides of the line the
   EA
   order parameter is  $q_1=p_1=0$. It is a transition in the liquid phase between 
   a regime of strong correlations
   between the two copies (large $\pt$) when $\eps$ is larger than the transition value,
   and a regime of smaller correlations, we shall
   call it the correlation transition. Along this line the entropy is discontinuous. 
   However these two transition lines (full line and dotted line)
   are not the analytic continuation one from the other:
   this is clearly seen from fig. \ref{entrop_jump} which plots the jump of entropy
   along the two discontinuous transition lines.
      The latent  heat is continuous across the triple point when
   one follows the discontinuous transition lines, but its derivative is discontinuous
   across the triple point. The 
   first order correlation transition ends up in the liquid phase  at a critical
   point where
   the correlation transition becomes of second order. This correlation transition
   is very similar to the familiar liquid-gas transition.
 
 It is interesting to notice how the glass transition is transmuted by going from
 a pure system to a system with two copies. In the pure system it is a
 'discontinuous spin glass' transition which is difficult to identify, being
 only signalled by a jump of the EA order parameter. When we switch on
  a small coupling $\eps$ between the copies, we get a real first order
  transition with an easily measurable jump in $\pt$ and a specific heat.
 This was already noticed by Franz and Parisi in their
 study of this same problem, as well as in their study of 
 two constrained systems, one being thermalised in the glass phase
 \cite{FraPar}. 
One outcome of the 
 present computation is that the glass transition is present at {\it all } $\eps$:
 there is no way one can go continuously from the high temperature phase to the glass phase
 without crossing a phase transition. This transition can be either 
 of the 'discontinuous spin glass' nature when the coupling $\eps$ is either zero
 or large, or it may be first order when the coupling is small enough and non zero, 
 because it is then accompanied by a correlation transition.

\subsection{Studying the glass through the cloned liquid}
\begin{figure}
\centerline{\hbox{
\epsfig{figure=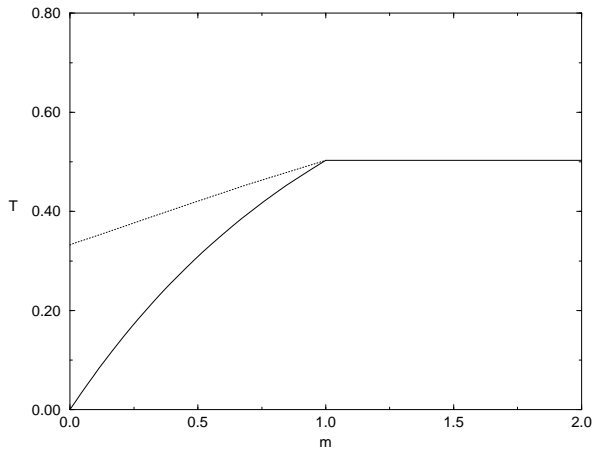,width=7 cm,angle=-90}
}}
\caption{ Phase diagram of the cloned
 $p$-spin glass with p=4 at small $\eps$ in the plane $m$ (number of clones) - temperature.
 The diagram is obtained from maximising (if $m<1$)
or minimizing the free energy of $m$ clones, with the high temperature-liquid free energy 
involving only the correlation $q$ between the clones. 
Above the full line the system is 
in the liquid phase, below it is a glass. The free energy along the
transition line at $m<1$ can be obtained from a high temperature computation in
a correlated liquid. As the free energy in the glass phase is independent of
$m$, this allows to compute the free energy in the glass phase.
 The dotted line is a first order transition 
line between a
liquid which is coupled, with a finite value of $q$,
 below the dotted line, and one where
the clones are uncoupled, ($q$), above the line}
\label{fig13}
\end{figure}

We now move to the case in which we have $m$ coupled clones, and study specifically
the case $m<1$. The general solution could be studied from the formulas derived
in \cite{kurparvir}, but here we want to concentrate on the test of the 
general method described above, which states that the thermodynamics of the glass 
can be studied from a high temperature computation of a correlated system. We thus
consider the $m$ systems with a fixed correlation between any of the 
$m(m-1)/2 $ pairs:
\be
{1 \over N} \sum_i S_i^a S_i^b =q
\ee
and compute the partition function
\bea
Z= \int \prod_{a=1}^m \((\prod_i d S_i^a \delta\((\sum_i( S_i^a)^2-N\))\))
\prod_{a<b} \delta \(({1 \over N} \sum_i S_i^a S_i^b -q \))
\\
\exp\((- \sum_{i_1<...<i_p} J_{i_1...i_p} \sum_{a=1}^m \ S_{i_1}^a...S_{i_p}^a \))
\eea
As we perform only a high temperature computation, we do not need to introduce
the  replicas in order to perform the 
quenched average. The simplest method is to notice that the annealed
average is exact in
the high temperature phase (or we can also expand $Z$
 in a high temperature series). 
 We find a free energy $ \phi=-\log Z/(\beta N m)$ equal to:
\be
\phi=-{\beta \over 4}[1-(1-m) q^p] + {1-m \over 2 \beta m} \log(1-q)
-{1 \over 2 \beta m} \log[1-(1-m) q]
\ee
This is exactly the free energy of a single, uncloned system, when it is studied
through the introduction of  replicas in (\ref{free_noclone}). 
Our number of clones $m$ is equivalent
to the breakpoint $x$ in the one step rsb Ansatz, and the correlation $q$ is 
equivalent to 
 the matrix element $q_1$ in the replica method. As explained in the previous section,
this could be 
expected on general grounds \cite{remi}. It also implies that 
the proposed method of the high temperature expansion of a cloned system
works. In fact if one studies, for $m<1$, the free energy $\phi$ using the general method 
outlined in sect. \ref{constrained}, one gets the phase diagram in the plane
$m,T$ shown in fig. \ref{fig13}. The value of the critical temperature
$T_c(m)$ is the inverse function of the $x(T)$ curved obtained with the replica solution,
and the thermodynamics deduced from sect. \ref{constrained} is automatically correct.

The interesting point of this study is the structure of the phase diagram
of fig. \ref{fig13}. It confirms our
expectations from the above  general discussion: the transition is
rather different depending on whether $m$ is larger or smaller than one. When $m<1$
the glass transition is between a correlated liquid and a glass. Therefore
the high temperature phase that must be studied in order to compute
the properties of the glass is that of the correlated liquid \cite{fn3}.  When $m>1$
the transition is between an uncorrelated liquid and a glass. This explains some
of the previous failures and successes of recent works on glasses using
this clonation idea. In \cite{MPhnc} the transition in soft or hard 
sphere systems was studied as one from an
uncorrelated liquid to a glass. This allowed to get a good result for the transition
temperature, but it did not give the correct glass
thermodynamics, because this glass thermodynamics should be studied from a
study of a correlated liquid, not from an uncorrelated one. 
The correct procedure was performed
 in \cite{MPglass} through a study of a molecular liquid.  

\section{Conclusion}
We have tested the general method of \cite{MPglass} on the solvable case of the $p$-spin model.
It works well in the sense that the study of a 'molecular liquid' with $m<1$ clones
allows to obtain the exact result for
the thermodynamics of the glass phase. Obviously the computations which are performed 
with the clones in their high temperature phase are equivalent to the direct 'one
step replica symmetry breaking' solution of the problem: the clones are
basically equivalent to replicas \cite{oKane,remi}. However one should not forget
that the clones give us a method to compute the thermodynamic properties 
of glasses even in systems without
any quenched-in disorder, where there is no way to introduce the replicas. 
In fact we believe that the method of replica symmetry breaking, although it was
historically introduced in order to compute some quenched averages, involving the logarithm
of a partition function, is much more general. At the one step level it is able to address any
kind of problem (with or without disorder) where the low lying minima are uncorrelated,
with a distribution which is unbounded but decreases faster than any power law
\cite{BouMez_extr}. This is very interesting since these problems build up one natural
universality class for extreme events statistics. The question of whether a given problem 
(e.g. a given type of structural glass) falls into this universality class is a difficult one.
If it does not, it means that there will be another phase transition which forbids the direct
study of the glass from that of the cloned liquid (a system with two step breaking could be
studied by cloning the clones...). This is very probably what happens in the 
Edwards Anderson model \cite{MPRR}. The case of structural glasses, if it is confirmed
that it belongs to this universality class of systems which are solved 
by cloning, (the favourable indications are reviewed in \cite{MPglass}), is thus
a much simpler one. Yet it would also be interesting to try to carry out 
a similar method on finite dimensional spin glasses, in order to see the predictions e.g. for
the ground state energy.

\section{Acknowledgements}
It is a pleasure to thank G.Parisi for many discussions and comments. This 
research was supported in part  by the National Science Foundation
under Grant No. PHY94-07194.

\end{document}